\documentclass{svmult}

\usepackage[sort&compress,numbers]{natbib}
\usepackage{amsmath}
\usepackage{type1cm} 
\usepackage{makeidx} 
\usepackage{graphicx}
\usepackage{xcolor}
\usepackage{multicol} 
\usepackage[bottom]{footmisc}

\usepackage{siunitx}
\usepackage{url}

\usepackage{newtxtext} 
\usepackage[varvw]{newtxmath} 
\usepackage[colorlinks=true,citecolor=blue,urlcolor=red]{hyperref}

\makeindex 

\begin{document}

\title*{Virtual potential created by a feedback loop: taming the feedback demon to explore stochastic thermodynamics of underdamped systems}
\titlerunning{Taming the demon behind a virtual feedback potential}
\author{Salambô Dago\orcidID{0000-0002-1964-2375},\\
Nicolas Barros\orcidID{0009-0000-1348-7725},\\
Jorge Pereda\orcidID{0000-0002-4802-0591},\\
Sergio Ciliberto\orcidID{0000-0002-4366-6094} and\\
Ludovic Bellon\orcidID{0000-0002-2499-8106}}
\authorrunning{S. Dago, N. Barros, J. Pereda, S. Ciliberto and L. Bellon}
\institute{University of Lyon, ENS de Lyon, CNRS, laboratoire de Physique, \email{ludovic.bellon@ens-lyon.fr}}

\maketitle

\abstract{Virtual potentials are an elegant, precise and flexible tool to manipulate small systems and explore fundamental questions in stochastic thermodynamics. In particular double-well potentials have applications in information processing, such as the demonstration of Landauer's principle. In this chapter, we detail the implementation of a feedback loop for an underdamped system, in order to build a tunable virtual double-well potential. This feedback behaves as a demon acting on the system depending on the outcome of a continuously running measurement. It can thus modify the energy exchanges with the thermostat and create an out-of-equilibrium state. To create a bi-stable potential, the feedback consists only in switching an external force between two steady values when the measured position crosses a threshold. We show that a small delay of the feedback loop in the switches between the two wells results in a modified velocity distribution. The latter can be interpreted as a cooling of the kinetic temperature of the system. Using a fast digital feedback, we successfully address all experimental issues to create a virtual potential that is statistically indistinguishable from a physical one, with a tunable barrier height and energy step between the two wells.}

\abstract*{Virtual potentials are an elegant, precise and flexible tool to manipulate small systems and explore fundamental questions in stochastic thermodynamics. In particular double-well potentials have applications in information processing, such as the demonstration of Landauer's principle. In this chapter, we detail the implementation of a feedback loop for an underdamped system, in order to build a tunable virtual double-well potential. This feedback behaves as a demon acting on the system depending on the outcome of a continuously running measurement. It can thus modify the energy exchanges with the thermostat and create an out-of-equilibrium state. To create a bi-stable potential, the feedback consists only in switching an external force between two steady values when the measured position crosses a threshold. We show that a small delay of the feedback loop in the switches between the two wells results in a modified velocity distribution. The latter can be interpreted as a cooling of the kinetic temperature of the system. Using a fast digital feedback, we successfully address all experimental issues to create a virtual potential that is statistically indistinguishable from a physical one, with a tunable barrier height and energy step between the two wells.}
	
\section{Introduction} \label{sec:intro}

Feedback potentials are widely used to trap and manipulate Brownian particles in solution, and explore fundamental questions in non-equilibrium statistical mechanics of small systems~\cite{Gavrilov-2014,Gavrilov-2016,Cohen-2005,Jun-2012}. Indeed, by controlling an external force acting on a colloidal particle as a function of its measured position, one can create a virtual potential. This is a very powerful tool, more flexible~\cite{Albay-2020} than its physical counterparts consisting of potential forces created by optical or magnetic tweezers~\cite{Berut-2012,Berut-2015,Hong-2016,Martini-2016,Proesmans-2020}. Feedback potentials are used in particular to study Landauer's principle, by creating a double well and using the trapped particle as a memory~\cite{Jun-2014, Proesmans-2020}. Within the information processing framework, lowering the dissipation seems a promising path to reduce energy costs~\cite{Dago-2021, Dago-2022-PRL, Gieseler-2018, Gieseler-2015}. Working with virtual potentials within underdamped dynamics thus appears as a natural endeavor. Moreover, the underdamped regime offers new insights on a wide variety of fundamental questions tackling the connections between feedback and thermodynamics~\cite{Seifert-2012, Kim-2004,Granger-2011, Rosinberg-2015,Sagawa-2010,Lahiri-2012}.

Nevertheless, implementing virtual potentials in the underdamped regime is not an easy task, especially within the stochastic thermodynamics framework that requires to work at the energy scale of $k_BT_0$, where $k_B$ is the Boltzmann constant and $T_0$ the temperature of the thermal bath. Indeed, resonant systems are very sensitive to perturbations, noises or drifts. Moreover, the feedback update delay can have strong consequences on the coupling between the system and the thermal bath~\cite{Jun-2012, Rosinberg-2015}.

We describe in this chapter an electrostatic feedback designed to create virtual double-well potentials acting on a micro-cantilever, which serves as an underdamped mechanical oscillator. The feedback acts as a demon, which must be tamed in order to produce a ``perfect'' virtual potential. Specifically it should not perform nor harvest any work on the system, nor modify the heat exchanges with the thermostat. Although the detailed thermodynamic properties of the demon are not the subject of this study, we show how the non-idealities of the feedback influence the quality of the virtual potential. After this fine tuning, the system offers a high flexibility and precision, with excellent quality in terms of position measurement and force tuning. We are able to create clean, reliable and tunable double-well potentials which outperform those produced by optical and magnetic tweezers (either physical or virtual), and have the added advantage of being analytically tractable. This work therefore presents an unprecedented experimental tool to explore information thermodynamics, and in particular Landauer's principle in the underdamped regime. 

In the following, we detail the experimental challenges one faces to remove any bias introduced by the feedback loop. This study incorporates experimental and numerical simulation results, as well as a comprehensive theoretical model. The latter includes the unified and complete description of the switching time of the cantilever in the double-well potential: our expression tends towards Kramer's escape time in the high-energy barrier limit, but it also provides an adjusted model for barriers lower than the thermal energy, where Kramer's formula is no longer valid.

The chapter is organised as follows: we first present the experimental system and the principle of the feedback loop to create a virtual potential indistinguishable from an equivalent physical one (section \ref{sec:principle}). We then explore the non-idealities of an experimental implementation (section \ref{sec:implementation}). In particular, we analyse how measurement noise or delays in the switches between the wells result in a bias of the energy exchanges with the thermal bath, effectively warming or cooling the oscillator Brownian noise. From this comprehensive analysis, we describe the digital filtering process with anticipation we need to implement to mitigate imperfections. Lastly, section \ref{sec:Teff} describes how we can use the feedback loop as a demon to induce a controlled heat flow between the thermal bath and the system, driving it into a non-equilibrium steady state. We finally conclude on the potential of this approach to explore the demon behavior from the information prism.

\section{Virtual double well potential: principle} \label{sec:principle}

In this section we describe first the experimental set-up whose main component is a cantilever behaving like an harmonic oscillator. We then present how the feedback force acts on the system and how it is calibrated in order to perform quantitative measurements and to compare the experimental results to the theoretical ones. We finally describe how double well potentials can be created, and the amount of energy exchanged due to the feedback when crossing the barrier from one well to the other. 

\subsection{From a simple oscillator to a bistable system}
\begin{figure}[b]
	\sidecaption[b]
 \centering
	\includegraphics[height=5.8cm]{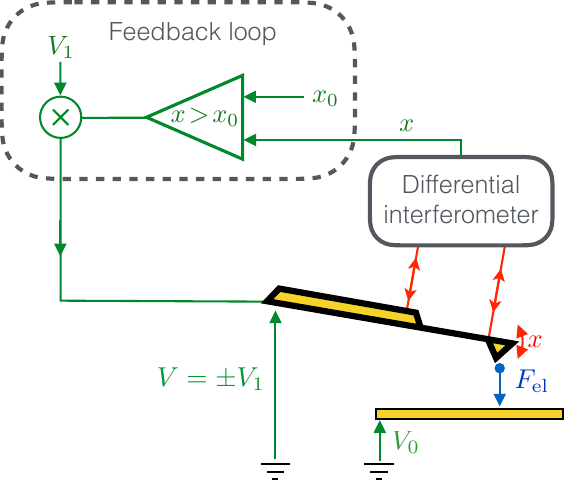}
	\caption{\textbf{Experimental system.} The conductive cantilever is sketched in yellow. Its deflection $x$ is measured with a differential interferometer~\cite{Paolino-2013}, by two laser beams focused on the cantilever and on its base. The cantilever at voltage $V=\pm V_1$ is facing an electrode at $V_0$. The voltage difference $V-V_0$ creates an attractive electrostatic force $F\propto(V-V_0)^2$. The double-well potential is created by the feedback loop, consisting in a comparator and a multiplier (dashed box).}
	\label{fig:chap2_schema_bloc}
\end{figure}

\begin{figure}[t]
	\sidecaption[t]
	\includegraphics{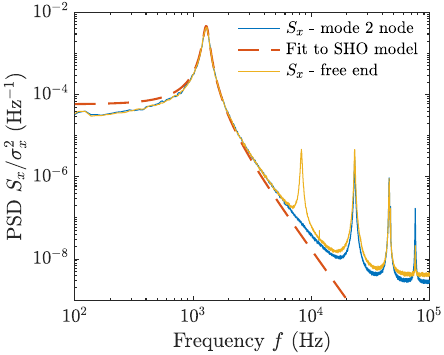}
	\caption{\textbf{PSD $S_x$ of the thermal noise driven deflection in a single well.} $S_x$ is measured with no feedback ($V_1=0$, solid lines), fitted by the theoretical spectrum of a SHO (dashed line). The second deflection mode, visible at $\SI{8}{kHz}$ when the laser beam is focused at the free end of the cantilever (yellow), is successfully hidden by focusing the laser beam on the node of this mode (blue). Up to $\SI{10}{kHz}$, the cantilever behaves like a SHO at $f_0=\SI{1270}{Hz}$, with a quality factor $Q=10$.}
	\label{FigPSD}
\end{figure}

As sketched in Fig.~\ref{fig:chap2_schema_bloc}, the underdamped oscillator is a conductive cantilever which is weakly damped by the surrounding air at room temperature $T_0$. Its deflection $x$ is measured with very high signal-to-noise ratio by a differential interferometer~\cite{Paolino-2013}. The Power Spectral Density (PSD) of the thermal fluctuations of $x$ is plotted in Fig.~\ref{FigPSD}: the fundamental mode dominates by 3 orders of magnitude the higher-order deflection modes of the cantilever. The second deflection mode at $\SI{8}{kHz}$ is conveniently removed from the measured signal by focusing the sensing laser beam on its node, at around $0.78\%$ of the cantilever length. This adjustment helps in having a physical system very close to an ideal Simple Harmonic Oscillator (SHO), characterized by its resonance frequency $f_0=\omega_0/2\pi$, mass $m$, stiffness $k=m\omega_0^2$ and quality factor $Q$. The fit of this PSD with the theoretical thermal noise spectrum of a SHO leads to $f_0=\SI{1270}{Hz}$, $k\sim\SI{4e-3}{N/m}$ and $Q\sim10$ in air. The slight difference between the measurement and the model is due to frequency dependence of the viscous damping of the cantilever in air~\cite{Sader-1998,Bellon-2008}. Higher quality factors can be achieved in vacuum. We used as the length scale the variance at equilibrium $\sigma_x^2 = k_B T_0/k \sim \SI{1}{nm^2}$.

As such, the cantilever is an underdamped oscillator moving in an harmonic potential of stiffness fixed by the spring constant of the mechanical beam. In order to use the cantilever in stochastic thermodynamics experiments, we need to be able to tune this potential. For example to use it as a one-bit memory and probe Landauer's bound~\cite{Dago-2021,Dago-2022-PRL,Dago-2023-PNAS}, we need to confine its motion in an energy potential consisting of two wells separated by a barrier, whose shape can be tuned at will. Physical potentials with non-linearities at the $k_BT_0$ scale are very difficult to create for such an object. We therefore turn towards a virtual potential $U$ created by a feedback loop.

In this chapter, we focus on the creation of a double well potential, and implement a feedback which compares the cantilever deflection $x$ to an adjustable threshold $x_0$. After having multiplied the output of the comparator by an adjustable voltage $V_1$, the result is a feedback signal $V=S(x-x_0)V_1$, with $S(\cdot)$ the sign function: $V=+V_1$ if $x>x_0$ and $V=-V_1$ if $x<x_0$. The voltage $V$ is applied to the cantilever which is at a distance $d$ from an electrode kept at a voltage $V_0$. The cantilever-electrode voltage difference $V_0\pm V_1$ creates an electrostatic attractive force $F=\frac{1}{2}\partial_dC(d)(V_0 \pm V_1)^2$~\cite{Butt-2005}, where $C(d)$ is the cantilever-electrode capacitance. Since $d\gg \sigma_x$, $\partial_dC(d)$ can be assumed constant. We apply $V_0\sim \SI{100}{V}$ and $V_1\ll V_0$ so that, to a good approximation, $F\propto \pm V_1$ up to a static term. This feedback loop results in the application of an external force whose sign depends on whether the cantilever is above or below the threshold $x_0$. If the feedback loop is fast enough, the switching transient is negligible. As a consequence, the oscillator evolves in a virtual static double-well potential, whose features are controlled by the two parameters $x_0$ and $V_1$. Specifically, the barrier position is set by $x_0$ and its height is controlled indirectly by $V_1$, which sets the wells centers $\pm x_1 = \pm V_1 \partial_dC(d)V_0/ k$. The potential energy constructed by this feedback is computed by integrating the total force (spring + electrostatic) as a function of position:
\begin{align}
U(x,x_0,x_1)=\frac{1}{2} k \big(x-S(x-x_0)x_1\big)^2 + kx_0x_1\big(S(x-x_0) +S(x_0)\big).
\label{eq_U(x,x0,x1)}
\end{align}

\begin{figure}[b]
	\sidecaption
	\includegraphics{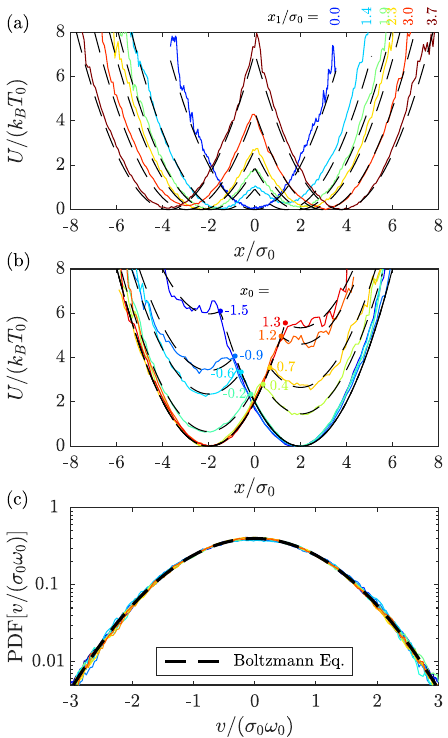}
	\caption{\textbf{(a)Virtual symmetric doubled well potentials} inferred from the position PDF and the Boltzmann equilibrium distribution, for half-distances $x_1=0$ to $3.7\sigma_x$ and a barrier position at $x_0=0$. Dashed line are best fits to Eq. \ref{eq_U(x,x0,x1)}. \textbf{(b) Virtual asymmetric doubled well potentials} for a half-distance $x_1=2\sigma_x$ and barrier positions from $x_0=-1.5$ to $1.3\sigma_x$. Dashed line are best fits to Eq. \ref{eq_U(x,x0,x1)}. \textbf{(c) Velocity PDFs} corresponding to the data of (a): they are indistinguishable from the one corresponding to the equilibrium in a single well. All the data presented in this figure is acquired in vacuum with a quality factor $Q=80$.}
	\label{FigDoubleWell}
\end{figure}

\subsection{The virtual double well potentials}

The potential energy in Eq.~\eqref{eq_U(x,x0,x1)} can be experimentally measured from the Probability Distribution Function $\mathrm{PDF}(x)$ and the Boltzmann equilibrium distribution: $\mathrm{PDF}(x)\propto e^{-U(x)/k_BT_0}$. Fig.~\ref{FigDoubleWell} presents examples of experimental (a) symmetric and (b) asymmetric double-well potentials generated by the feedback loop we describe in this chapter. The dashed lines are the best fits with Eq.~\eqref{eq_U(x,x0,x1)}, demonstrating that the feedback-generated potential behaves as a static one, in terms of the position PDF.

The second degree of freedom of the underdamped system is the velocity $v=\dot x$, and it is also expected to satisfy the Boltzmann equilibrium distribution. The equilibrium PDF of the velocity in the double-well should be the same as the one in a single harmonic well: a Gaussian of variance $\sigma_v^2=k_BT_0/m$ $\mathrm{PDF}(v)\propto e^{-mv^2/(2k_BT_0)}$. As shown in Fig.~\ref{FigDoubleWell}(c), our implementation of the feedback also validates this point.

In the following, we present the strategy we implemented to create such a virtual potential. Using such a fine tuned feedback loop, or good behaving demon, the potential is identical to a physical one, without any noticeable effect on the position and velocity equilibrium distributions.

\subsection{Energy exchanges when switching wells}

The main principle behind the virtual potential construction is to use a demon that switches the harmonic well center exactly when $x$ crosses the threshold $x_0$. Let us analyse here the upward switch from $-x_1$ to $+x_1$, the extension to the other case is straightforward. If the demon acts right on time with a negligible switching time, then the oscillator evolves in the desired potential defined by Eq.~\ref{eq_U(x,x0,x1)}, and no energy is exchanged during the switch: the demon is invisible with respect to the defined potential.

\begin{figure}[b]
	\sidecaption[b]
	\includegraphics{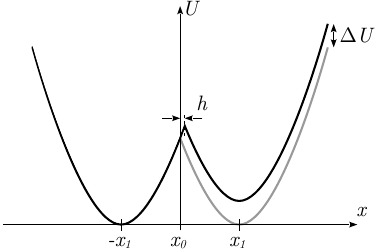}
	\caption{\textbf{Energy associated with an error on the threshold}: if instead of switching wells when $x$ crosses $x_0$, the demon acts at $x_0+h$ with $h>0$, the right well is a bit higher than expected, by $\Delta U = 2kx_1h$.}
	\label{FigDeltaU}
\end{figure}

However, the demon acts according to the available information, that is the measured position $x_m$, which can be noisy, or somewhat delayed with respect to the actual position of the oscillator $x$. It means that in practice, the demon switched when the oscillator crossed $x=x_0+h$, with $h$ a small quantity describing the error with respect to the desired threshold. Right after switching, the oscillator is thus evolving in the potential $U(x,x_0+h,x_1)$, which is shifted by $\Delta U (h) = 2kx_1h$ with respect to the goal potential (illustration Fig.~\ref{FigDeltaU}). This $\Delta U$ can be understood as the energy taken or given by the demon (depending on the sign of $h$) to the oscillator, supposed to be evolving in the potential $U(x,x_0,x_1)$. To further illustrate this point, let us consider the kinetic energy of the mechanical system, assuming a high quality factor, thus low energy exchanges with the thermostat on short time scale. If $h>0$, the oscillator had to climb the initial potential a bit higher than anticipated before switching to the next well, thus loosing the kinetic energy equivalent to $U(x_0+h,x_0+h,x_1)-U(x_0+h,x_0,x_1)=2kx_1h$: the oscillator is somewhat colder than it should have, this energy was taken by the demon from the system. If $h<0$, the situation is opposite and the demon gives energy to the oscillator which get somewhat hotter.

When analyzing the downward switch from $+x_1$ to $-x_1$, the conclusion are equivalent, up to a sign reversal for $h$: $\Delta U (h) = -2kx_1h$. $h>0$ is anticipating the switch and warms the system, while $h<0$ is delaying the switch and pumps energy from the system. Upon multiple crossing of the barrier, all these energy exchanges add up and can create a net energy flux to or from the system to the demon. If $h$ is a deterministic or random variable independent of the direction, then in average the net flux is zero and the demon is transparent: the effective mean potential is indistinguishable from a physical one.

\section{Digital feedback loop implementation} \label{sec:implementation}

In this section we describe the task of performing the switch between wells at the right time. Noise creates early and repeated switches which should be avoided (wrong virtual potential, warming), so we need to filter the high frequency noise. This process induces a delay in the response, resulting in an effective alternating asymmetric double wells which in turns cools the system. So we need to anticipate the signal to compensate for this delay and finally reach our goal. After describing this strategy, we present how to implement it in a digital feedback loop, going into the details of the technical steps of the filter, and of the resulting transfer function.

\subsection{Taming the demon} \label{sec:tamingthedemon}

The task of the demon inside the feedback loop is to switch the harmonic well center to the right position $\pm x_1$ whenever the position signal $x$ crosses the threshold $x_0$. As simple as this operation may look like, the real world implementation is not straightforward. Indeed, as illustrated by the power spectral density of the signal plotted in Fig~\ref{FigPSD}, on top of the signal of interest $x$ (the first mode of the oscillator), high frequency noise is present in the measured signal $x_m$. This noise originates from higher order resonance modes (mode 3 and above, thanks to the trick of canceling the second mode with the proper choice of the probe laser) of the cantilever, as well as the floor noise of the interferometer due to the shot noise of the photodiodes.

\begin{figure}
	\sidecaption[t]
	\includegraphics{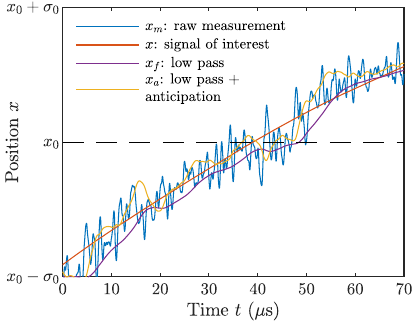}
	\caption{\textbf{Triggering right on the threshold}: the measured signal $x_m$ is hampered by high frequency noise with respect to the signal of interest $x$. This noise creates detrimental multiple and early triggering. Standard low pass filtering delays the signals and thus triggers late. We implement a low pass filter with an anticipation based on the speed to detect the crossing of $x_0$ on time in average.}
	\label{FigFilters}
\end{figure}

As illustrated in Fig.~\ref{FigFilters} on synthetic signals, this high frequency noise has an important consequence on the comparator: repeated high frequency commutation between the two wells. This effect is counterproductive in our case: during fast and repeated switches between $\pm V_1$, the average voltage experienced by the cantilever is typically $0$, therefore the mode 1 of the oscillator responding only to low frequencies can be stabilized in a well centered around the origin. This effect is generally canceled in standard comparators by introducing an hysteresis $h>0$ larger than the noise amplitude: once triggered ($x_m>x_0+h$ for instance), $x_m$ is sufficiently far from the new threshold $x_0-h$ to trigger the comparator again. This approach is however forbidden in our case: it would result in a systematic pumping of energy $\Delta U = 2kx_1h >0$ at each barrier crossing in both directions, thus effectively cooling the system. Such a demon would result in a constant heat flow from the oscillator, creating a non-equilibrium steady state and modifying the distribution of velocities.

As an alternative, we implement a temporal lock-up to freeze the comparator state after a switch, for $1/4$ of the oscillator's natural period $1/f_0$. By the time the comparator is active again, the cantilever has evolved in the new well on average long enough to reach the bottom of the well, and is therefore far enough from the threshold that an undue noise-induced switch is improbable. One drawback is that short excursions in the other well are forbidden as well. However these events, indeed present in a real double-well potential, are unlikely enough that removing them has no noticeable effect of the statistical properties of the virtual potential.

This temporal lock-up strategy only partially solves the problem of high frequency noise. Indeed, this noise results in an early trigger, since the measured signal $x_m$ always presents a fluctuation crossing the threshold before $x$ does. It corresponds to an effective negative hysteresis $h<0$, which results in a systematic injection of energy $\Delta U = 2kx_1h <0$ when crossing the barrier in both directions, thus effectively heating the system. For example, the thermal noise amplitude of mode 3 alone is $\num{4e-2}\sigma_x$, corresponding to $\Delta U \sim -0.08 k_BT_0$ at each crossing for a barrier height $\mathcal{B}=\frac{1}{2}{k_BT_0}$. Again, such a demon would result in a non-equilibrium steady state with a constant heat flow to the oscillator.

We therefore need to filter these high frequency fluctuations, using a low pass filter cutoff frequency $f_c$ below mode 3 resonance $f_3=\SI{23}{kHz}$, choosing for example $f_c=\SI{10}{kHz}$. However, this filtering induces a phase delay $\phi$ via its transfer function. For the third order Butterworth filter that we use, this phase is $\phi_\mathrm{BW_3} \sim 2 f/f_c \sim0.1$ in the frequency range of interest, close to the resonance of mode 1. The filtered signal $x_f$ is therefore delayed by $\tau_d \sim 1/(\pi f_c)=\SI{32}{\mu s}$ with respect to $x$. Since the typical speed when crossing the threshold is $\sigma_v=\sigma_x \omega_0$, the delay corresponds to an average positive hysteresis $h=\sigma_v\tau_d=2 \sigma_x f_0/f_c \sim 0.2 \sigma_x$, thus to $\Delta U \sim 0.4 k_BT_0$ at each crossing for $x_1=\sigma_x$. One more time, such a demon would result in a constant heat flow from the oscillator, creating a non-equilibrium steady state.

However, since this delay $\tau_d$ due to the filter is known, one can measure the actual speed of the oscillator and anticipate when the value of $x$ will cross the threshold, and thus trigger in average right on time. This strategy is illustrated in Fig.~\ref{FigFilters}: the speed, computed on the filtered signal $x_f(t)$, is use to construct $x_a(t)=x_f(t)+\tau_a \dot x_f(t) \approx x_f(t+\tau_a)$. Since $x_f(t) \approx x(t-\tau_d)$, $x_a(t) \approx x(t-\tau_d+\tau_a)$ can be a good forecast of the signal of interest choosing $\tau_a=\tau_d$. This is the final strategy that we implement in our feedback loop, and we adjust the value of $\tau_a$ of the experimental setup to compensate both for filtering and additional delays induced by the total feedback loop. Such taming of our demon results in the data displayed in Fig.~\ref{FigDoubleWell}, where the virtual potential is indiscernible from a physical one, both in position and speed.

\subsection{Practical implementation}

To allow for fast and versatile strategies, our feedback loop is implemented using a FPGA (field-programmable gate array) board associated with fast analog to digital (ADC) and digital to analog (DAC) circuits. A Labview-based FPGA data acquisition system from NI (FPGA 7975R + Adapter module NI 5783) allows operations clocked at $\SI{10}{ns}$ (\SI{100}{MHz} acquisition frequency, \SI{200}{MHz} onboard clock), with the ADC and DAC operations (16 bits) performed in approximately $\SI{500}{ns}$. The onboard operations are the following:
\begin{enumerate}
 \item ADC: sampling of the 4 photodiodes signal of the interferometer. Those signals of amplitude around $\SI{0.5}{V}$ are collected at the output of analog current to voltage amplifiers with a gain of $\SI{e5}{V/A}$ and a bandwidth of $\SI{1}{MHz}$~\cite{Paolino-2013}.
 \item Computing of the deflection $x_m$ of the cantilever. This step implies an arctan function, and multiplication by pre-calibrated coefficients, and last $\SI{270}{ns}$ (27 clock ticks).
 \item Digital filtering. Using a single digital IIR filter (infinite impulse response) combining a third order Butterworth filter and an anticipation by $\tau_a$ of the position, we compute $x_a$. This steps is $\SI{100}{ns}$ long (10 clock ticks).
 \item Comparison to the threshold. $x_0$ is extracted from a lookup table (to allow time dependent protocols), and compared to $x_a$. According to the result, a voltage $V_1$ or $-V_1$ is selected from a second lookup table.
 \item DAC: the voltage $V_1$ is sent to the device output.
 \item Export data: in parallel to the feedback operations, $x_m$, $x_0$ and the output voltage $V=\pm V_1$ are low pass filtered (antialiasing cutoff frequency $\SI{670}{kHz}$ of an independent IIR filter), down-sampled at $\SI{2}{MHz}$, and exported to the host computer for post-processing.
\end{enumerate}
The total deterministic delay to be anticipated by these operations is $\tau_d\sim\SI{1}{\mu s}$. Some details of each steps are described in the next paragraphs

\subsubsection{Before the experiment: calibration}
We first perform a calibration step implementing in the FPGA the calibration coefficients to convert the interferometer four photodiodes voltage signals to an actual position $x$ in nm~\cite{Paolino-2013}. We then perform a ramp in the voltage $V$ applied to the tip and read the cantilever average position $\langle x_m \rangle$, to convert $V_1$ into $x_1$ and vice-versa. The origin of the $x$ axis is periodically (every protocol) set to $(\langle x \rangle =0, V=0)$ in order to remove the drift in position during long experiments.
 
\subsubsection{Filtering}

The raw signal $x_m$ should be low pass filtered ($x_f=\mathrm{BW_3}\otimes x_m$, with $\mathrm{BW_3}$ a third order Butterworth filter of low pass cutoff frequency $f_c=\SI{10}{kHz}$) and anticipated ($x_a=x_f+\tau_a \dot x_f$) before being compared to the threshold $x_0$. In the frequency space, those operations correspond to the following response function:
\begin{eqnarray} 
 x_a(\omega) & = & (1+i\omega\tau_a)x_f(\omega)=\frac{1+i\omega\tau_a}{(1+i\omega\tau_c)(1+i\omega\tau_c-\omega^2\tau_c^2)}x_m(\omega)\\
 & = & \frac{\sum_{k=0}^3 d_k (i\omega\tau_c)^k}{\sum_{k=0}^3 c_k (i\omega\tau_c)^k} x_m(\omega) \label{EqFilterFreqSpace}
\end{eqnarray}
where $\tau_c=1/(2\pi f_c)$, $C=[c_0,c_1,c_2,c_3]=[1,2,2,1]$ and $D=[d_0,d_1,d_2,d_3]=[1,\tau_a/\tau_c,0,0]$. In the discrete time domain where the FPGA operates, this filter should be translated in operations on the sampled values $x_m(t_n)$ at each time sampling $t_n=n\delta t$, with $\delta t=\SI{10}{ns}$. We implement this filter using an IIR scheme: the filter output at time $t_n$ is computed from the previous value of the output and of the input, as:
\begin{equation} \label{EqFilterIR}
 x_a(t_n)=\frac{1}{a_0}\left(\sum_{k=0}^3 b_k x_m(t_{n-k})-\sum_{k=1}^3 a_k x_a(t_{n-k})\right)
\end{equation}
where the coefficients $A=[a_k]_{k=0\ \mathrm{to}\ 3}$ and $B=[b_k]_{k=0\ \mathrm{to}\ 3}$ have to be inferred from the desired frequency response of Eq.~\ref{EqFilterFreqSpace}. Noting that $i\omega x(\omega)$ in the frequency domain corresponds to the discrete derivative $[x(t_n)-x(t_{n-1})]/\delta t$ in the discrete time domain, we construct the translation from one domain to the other as $A=MC$ and $B=MD$, where the matrix $M$ is defined by 
\begin{equation}
 M=\left[\begin{array}{rrrr}
 1 & \alpha & \alpha^2 & \alpha^3 \\
 0 & \ \ -\alpha & \ -2\alpha^2 & \ -3\alpha^3 \\
 0 & 0 & \alpha^2 & 3\alpha^3 \\
 0 & 0 & 0^{\color{white}2} & -\alpha^3
 \end{array}\right]
\end{equation}
with $\alpha=\tau_c/\delta t$. From the desired frequency response (coefficients $C$ and $D$), one can compute the IIR coefficients $A$ and $B$ to implement in the FPGA programming.

When following this strategy, we run into two sources of instabilities for the IIR filters. The first one is due to the the very low cutoff frequency $f_c=\SI{10}{kHz}$ with respect to the sampling frequency $f_s=\SI{100}{MHz}$: these 4 orders of magnitude makes the operation of Eq.~\ref{EqFilterIR} very sensitive to rounding errors, has we compute $x_a(t_n)$ from the difference of very large numbers compared to the result. We therefore need to perform 64 bits operations during the filtering process. The increased computational cost translates into a second source of instability of the IIR filter: each multiplication in Eq.~\ref{EqFilterIR} requires 9 FPGA ticks. As such, at time $t_n$ the information available is $t_{n-9}$ and not $t_{n-1}$. We therefore need to rewrite Eq.~\ref{EqFilterIR} as:
\begin{equation} \label{EqFilterIIR}
 x_a(t_n)=\frac{1}{a_0'}\left(\sum_{k=0}^3 b_k' \bar{x_m}(t_{n-9k})-\sum_{k=1}^3 a_k' x_a(t_{n-9k})\right)
\end{equation}
where the coefficients $A'=[a_k']_{k=0\ \mathrm{to}\ 3}$ and $B'=[b_k']_{k=0\ \mathrm{to}\ 3}$ are now computed with $\alpha'=\tau_c/(9\delta t)$. To avoid running 9 independent IIR filters in parallel, we actually use as an input the average signal $\bar{x_m}$ obtained with a sliding rectangular window on 9 samples:
\begin{equation}
 \bar{x_m}(t_n)=\frac{1}{9}\sum_{k=1}^{9}x_m(t_{n-k}).
\end{equation}
Formally, this is a low pass FIR (finite impulse response) filter applied to $x_m$ before the IIR filter, with a cutoff frequency of $f_s/9=\SI{11}{MHz}$.

\begin{figure}
	\centering
	\includegraphics[width=\textwidth]{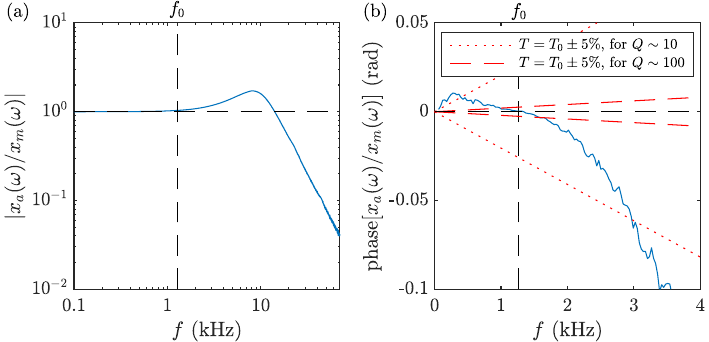}
	\caption{\textbf{(a) Gain of the FPGA position signal conditioning}. As expected the gain of the feedback operations on the position signal (ADC, conversion of the photodiode outputs into $x$ signal, IIR filter and DAC) is $1$ (horizontal dashed black) within the frequency range of interest ($f_0$ in vertical dashed black), and is then filtered above $f_{c}=\SI{10}{kHz}$ to remove the 3rd deflection mode contribution. We notice that there is a slight amplification above $f_0$ due to the resonance of the IIR filter. \textbf{(b) Phase of the FPGA position signal conditioning}: the IIR filter is designed to have a 0 phase for the overall processing of the position signal at $f_0$ (vertical dashed black), to avoid switching advance or delay, and hence meet the $T=T_0\pm5\%$ requirement on the kinetic temperature. The authorized range (computed with Eq.~\ref{EqPhaseCriterium}) corresponds to the interval between the red dashed lines for $Q\sim100$, and between the dotted lines for $Q\sim10$.}
	\label{FPGABode}
\end{figure}

Implementing all those bricks results in a stable and efficient digital filter, designed to remove high frequency fluctuations and correct for the filter dephasing and the $\sim\SI{1}{\mu s}$ feedback delay. Formally, we would like the overall phase of FPGA feedback to vanish around the oscillator's resonance. In practice, the anticipation coefficient $\tau_a$ is fine tuned experimentally to enforce an unbiased velocity PDF in the resulting double-well potential. The Bode diagram of the overall FPGA operations on the position signal (ADC, conversion of the photodiode outputs into $x_m$ signal, FIR and IIR filter, DAC) is displayed in Fig.~\ref{FPGABode}. It consists in the gain and the phase of the transfer function between the real time position directly inferred from the photodiode outputs (no real time conversion, nor filtering, but only post treatment operations), and the position signal acquired, reconstructed, filtered and output by the FPGA device, $x_a$. It is worth noticing on Fig.~\ref{FPGABode}(a) that this real time zero-phase -at $f_0$- filtering is performed at the expense of a very small resonance in the filter gain, which has negligible consequences on our experiments.

\subsection{Recording and post-treatment analysis}
 
To compute the thermodynamics quantities we use the measured position $x_{\textrm{acq}}$ properly filtered above $\SI{10}{kHz}$ in post-treatment analysis using a zero-phase filtering (processing the data in both the forward and reverse direction). This post-acquisition filter is better than the real time FPGA one (because it enforces 0 phase for all frequencies and not only $f_0$), that is why we do not use directly $x_f$ for the analysis. Let us note that such zero phase filtering is not causal and thus cannot be implemented on the FPGA for real time operations. Thanks to the calibration of the data, no further post-treatment is required.

\section{Creating non-equilibrium steady states with unleashed demons} \label{sec:Teff}

In this section we explore the properties of the steady state when the total feedback delay is not tuned to zero: each time we cross the barrier, we lose or gain energy. Therefore we need to compute the barrier crossing rate, and from this we can deduce the effective kinetic temperature of the system, that we compare to the experiment.

\begin{figure}[t]
\sidecaption[t]
\includegraphics[width=7.5cm]{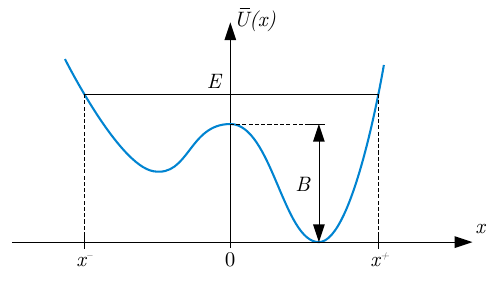}
\caption{We study the escape rate from one well ($x>0$) to the other $(x<0)$ on a generic double well potential $\bar{U}(x)$. The barrier height is $B$, and is positioned at $x=0$. For a total energy $E>B$ of an Hamiltonian system evolving in this potential, the motion is periodic between $x^-$ and $x^+$. For $E<B$, the system is stuck in the well of its initial condition.}
\label{fig-U}
\end{figure}

\subsection{Beyond Kramer's escape rate: switching between wells separated by a low barrier}

Let us study a generic one degree of freedom mechanical system of mass $m$ and position $x$ evolving in a potential energy $\bar{U}(x)$. Its kinetic energy is $K=p^2/(2m)$, with $p=m dx/dt$ the momentum, so that its Hamiltonian is:
\begin{equation}
H(x,p)= \frac{1}{2m}p^2+\bar{U}(x).
\end{equation}
We suppose it is statistically at temperature $T$: the initial conditions in position and momentum are drawn from a Boltzmann distribution
\begin{align}
P(x,p)&=\frac{1}{Z}e^{-\beta H(x,p)}, \label{eq-Boltzmanxp}\\
Z&=\iint dx \,dp \,e^{-\beta H(x,p)},
\end{align}
with $\beta=1/(k_B T)$.

We suppose that the potential presents a double well, as sketched in Fig.~\ref{fig-U}. The barrier between the two wells is placed in $x=0$ and its energy is $\bar{U}(0)=B$, with the origin of $\bar{U}$ at its minimum. In the limit of weak damping, the total energy $E=H(x,y)$ of the system is conserved in time. Therefore, its Hamiltonian dynamics can be solved for a given total energy $E$ by integrating the following equation:
\begin{equation}
\frac{dx}{dt} = \pm \sqrt{\frac{2}{m}[E-\bar{U}(x)]}.
\end{equation}

The motion is periodic: the time $\bar{\mathcal{T}}(E)$ to explore the full phase space available for a given $E$, depends on the value of $E$ with respect to the barrier height $\mathcal{B}$:
\begin{itemize}
\item If $E>\mathcal{B}$ the system explores both wells every period, so there are 2 barrier crossing (one in each direction) every period, with the period being:
\begin{align}
\bar{\mathcal{T}}(E>\mathcal{B}) &= 2\int_{x^-}^{x^+} \frac{dx}{\sqrt{2[E-\bar{U}(x)]/m}},
\end{align}
where $x^-$ and $x^+$ are the minimum and maximum values of $x$ allowed by the criterion $\bar{U}=E-K\leq E$.
\item If $E<\mathcal{B}$, then the motion is confined to a single well, there are no switches, there are actually 2 periods $\bar{\mathcal{T}}^-$ and $\bar{\mathcal{T}}^+$ to consider depending on which well is explored given the initial conditions. 
\end{itemize}
We define in the following the time spent in each well on a period as:
\begin{align}
\bar{\mathcal{T}}^-(E)&=2\int_{x^-}^{0} \frac{dx}{\sqrt{2[E-\bar{U}(x)]/m}},\\
\bar{\mathcal{T}}^+(E)&=2\int_{0}^{x^+} \frac{dx}{\sqrt{2[E-\bar{U}(x)]/m}},
\end{align}
where again the bound $0$ of the integrals should be adjusted when $E<\mathcal{B}$. When $E>\mathcal{B}$, we trivially verify $\bar{\mathcal{T}}(E)=\bar{\mathcal{T}}^-(E)+\bar{\mathcal{T}}^+(E)$, and we extend this relation to $E<\mathcal{B}$ to define the time to explore the full space available for a given energy.

As an illustration, we can use the symmetric bi-quadratic potential $U=\frac{1}{2} k (|x|-x_1)^2$, corresponding to $x_0=0$ and $\mathcal{B}=\frac{1}{2} k x_1^2$. We compute
\begin{subequations} \label{EqPeriodSymDblWell}
\begin{eqnarray}
 \mathrm{for}\ E<\mathcal{B}: & \frac{1}{2}\mathcal{T}(E) = \mathcal{T}^+(E) = \mathcal{T}^-(E) = & \mathcal{T}_0,\\
 \mathrm{for}\ E>\mathcal{B}: & \frac{1}{2}\mathcal{T}(E) = \mathcal{T}^+(E) = \mathcal{T}^-(E) = & \mathcal{T}_0 \left[\frac{1}{2}+ \frac{1}{\pi}\sin^{-1}\left(\sqrt{\frac{B}{E}}\right)\right].
\end{eqnarray}
\end{subequations}
$\mathcal{T}(E)$ is twice the natural period $\mathcal{T}_0= 2\pi/\omega_0$ of a single oscillator when $E\leq \mathcal{B}$ or $E\gtrsim\mathcal{B}$, and tends to $\mathcal{T}_0$ for $E\gg\mathcal{B}$.

Since time is a natural variable for this problem, we operate a change of variables from $(x,p)$ to $(t,E)$: for any given position and momentum, we can define a unique time and energy, and vice-versa. We consider here that as the system is periodic, the time range is restricted from $-\bar{\mathcal{T}}^-(E)$ to $+\bar{\mathcal{T}}^+(E)$ for a given total energy $E$. This is a canonical transformation for the Hamiltonian since the Jacobian determinant is 1:
\begin{equation}
dx\,dp=\left| \frac{\partial x}{\partial E} \frac{\partial p}{\partial t} - \frac{\partial x}{\partial t} \frac{\partial p}{\partial E}\right| dt\,dE= dt \,dE,
\end{equation}
The change of variable is therefore straightforward for the probability: 
\begin{align}
P(t,E)&=\frac{1}{Z}e^{-\beta E}, \label{eq-BoltzmantE}
\end{align}
where $P(t,E)$ is defined only for $-\bar{\mathcal{T}}^-(E)<t<\bar{\mathcal{T}}^+(E)$. The probability of having an energy $E$ is deduced by integrating on time, leading to:
\begin{align}
P(E)&=\frac{1}{Z}e^{-\beta E}\bar{\mathcal{T}}(E),\\
\mathrm{with}\ Z&=\int_0^\infty dE \,e^{-\beta E}\bar{\mathcal{T}}(E).
\end{align}
This computation can be done while limiting the explored space to the left or right well, leading to the same expression where we only substitute $\bar{\mathcal{T}}(E)$ by $\bar{\mathcal{T}}^\pm(E)$: $P^\pm(E)=e^{-\beta E}\bar{\mathcal{T}}^\pm(E)/Z^\pm$, with $Z^\pm=\int dE \,e^{-\beta E}\bar{\mathcal{T}}^\pm(E)$.

We can now compute the escape rate from one well to the other, lets' say $\Gamma^+$ from $x>0$ to $x<0$. For a given energy $E$, the escape rate is 0 if $E<\mathcal{B}$, and $1/\bar{\mathcal{T}}^+(E)$ otherwise (the system spend one period in the well before switching to the other, so one crossing every period). The mean transition rate is therefore
\begin{align} \label{eq-Gamma+}
\Gamma^+&= \int_{\mathcal{B}}^{\infty} dE\, \frac{1}{\bar{\mathcal{T}}^+(E)} P^+(t,E) = \int_\mathcal{B}^\infty dE \, \frac{1}{\bar{\mathcal{T}}^+(E)} \frac{1}{Z^+} e^{-\beta E} \bar{\mathcal{T}}^+(E)\\
&= \frac{\int_\mathcal{B}^\infty dE\ e^{-\beta E} }{\int_0^\infty dE\ e^{-\beta E}\bar{\mathcal{T}}^+(E)} = \frac{e^{-\beta \mathcal{B}} }{\int_0^\infty dE \beta e^{-\beta E} \bar{\mathcal{T}}^+(E)}\label{eq-Gammadblwell}.
\end{align}
This expression is generic and applies to any double well potential shape. It depends only on the potential shape of the considered well: the other well details have no effect on the escape rate. It also depends on the temperature and barrier height only through $\beta\mathcal{B} = \mathcal{B}/(k_BT)$.

When the barrier $\mathcal{B}$ is high ($\beta\mathcal{B}\gg1$), then the denominator to Eq.~\ref{eq-Gammadblwell} can be seen as the average period in the bottom of the well. Since most of the time a quadratic approximation will hold at the thermal noise level, this denominator is simply a period $\bar{\mathcal{T}}^+(0)=\mathcal{T}_0=2\pi/\omega_0$ of this harmonic oscillator, and we get a very simple expression matching Kramers' expression:
\begin{equation}
\Gamma^+ = \frac{1}{\mathcal{T}_0}e^{-\beta \mathcal{B}}.
\end{equation}

Let us now study the switching rate in the double well potential, that is the rate at which the system crosses the barrier, regardless of the direction. For an energy $E>\mathcal{B}$, this will occur twice every $\mathcal{T}(E)$. The switching rate $\Gamma$ thus writes:
\begin{align}
\Gamma&= \iint_{E>\mathcal{B}} dE\,dt\, \frac{2}{\bar{\mathcal{T}}(E)} P(t,E)\\
&= \frac{2e^{-\beta \mathcal{B}} }{\int_0^\infty dE\ \beta e^{-\beta E} \bar{\mathcal{T}}(E)} \label{EqExtKramers}\\
&= \frac{2}{1/\Gamma^{+}+1/\Gamma^{-}}
\end{align}
For a symmetric double well, the switching rate is equal to the escape rate of a single well: $\Gamma=\Gamma^{+}=\Gamma^{-}$.

\begin{figure}[t]
\sidecaption[t]
\includegraphics{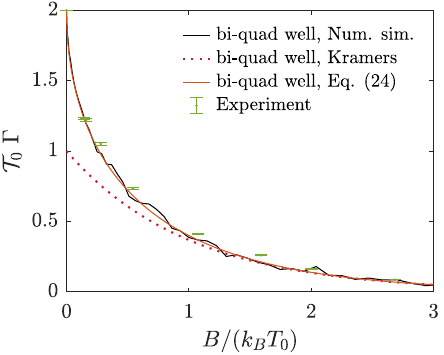}
\caption{{\bf Barrier crossing rate $\Gamma$} in a bi-quadratic well $U(x) = \mathcal{B} (|x|/x_1-1)^2$. The barrier height is $\mathcal{B}=\frac{1}{2} k x_1^2$, and $\mathcal{T}_0=2 \pi/\omega_0$ is the period in a single harmonic well. At low barrier height, Kramers' simple expression is a factor 2 below the result from Eq.~\ref{EqExtKramers}. The latter perfectly matches the experimental data or the results from a numerical simulation of a Langevin dynamics in the potential $U(x)$, using a quality factor $Q=10$.}
\label{fig:NumSimvsTh}
\end{figure}

To conclude this section, we apply the computation of the barrier crossing rate to the case of the bi-quadratic symmetric double well $U(x)$. For this, we simply report the value of $\mathcal{T}(E)$ from Eq.~\ref{EqPeriodSymDblWell} into Eq.~\ref{EqExtKramers}.
In Fig.~\ref{fig:NumSimvsTh}, we compare the barrier crossing rate computed from this last equation to the one measured on a numerical simulation of a Langevin dynamics in such a potential: the agreement is excellent. The Kramers approximation holds only in the limit of large barriers. It is interesting to note that though the derivation has been done in the Hamiltonian limit ($Q=\infty$), it is still valid even for moderate $Q$: using numerical simulation, we need to reach $Q=0.01$ to see a significant deviation to this model.

\subsection{Effective temperature due to the feedback delay}

Let us introduce the kinetic temperature $T$ of the oscillator defined through the velocity variance: $\sigma_v^2=\langle v^2 \rangle =k_B T/m$. At equilibrium in a bi-quadratic potential, the kinetic temperature should match the bath temperature $T_0$ as prescribed by the Boltzmann distribution. When out of equilibrium, we study the steady state and make the hypothesis that the velocity distribution is still described by a Gaussian. Thus the second moment quantified by $T$ caries all the information needed to describe the statistics of $v$.

\begin{figure}[htb]
 \begin{center}
 \includegraphics{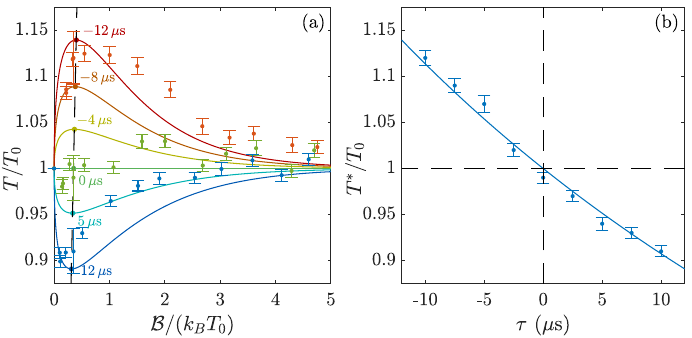}
 \end{center}
 \caption{\textbf{Consequences of a delay on the kinetic temperature.} (a) $T/ T_0$ is plotted as a function of the barrier height $\mathcal{B}$ between the wells, for various values of the feedback delay $\tau$ from $-\SI{12}{\mu s}$ to $+\SI{12}{\mu s}$. The continuous curves are computed with the model prediction provided by Eq.~\eqref{eqTdelay}, with $Q\omega_0=\SI{5.4e4}{rad/s}$. The dependence on $\mathcal{B}$ comes from the balance between the barrier crossing rate and the energy exchanges associated to the delay at each switch. The extremum temperature $T^*$ is well described by Eq.~\ref{EqT*} (black dashed line). The experimental data are recorded in air ($Q\sim10$), with $\tau_a$ adjusted to probe three different delays: $\tau=\SI{10}{\mu s}$ (late trigger), $\tau=\SI{0}{\mu s}$ (behaving demon), and $\tau=\SI{-10}{\mu s}$ (early trigger). When $\tau=0$, the kinetic temperature deviates at most by 3\% from the thermostat temperature. (b) Extremum temperature $T^*$ as a function of the feedback delay $\tau$, described by Eq.~\ref{eq:thetamin}. Positive delay $\tau$ cools the system down, while negative ones warms the system up. The experimental data corresponds to a measurement at $\mathcal{B}=b^* k_BT_0$, and is very close to the prediction.}
 \label{theta_map} 
\end{figure}

Let us study the non-equilibrium steady state of a wild demon, induced by a total time delay $\tau=\tau_d -\tau_a \neq 0$ between the barrier crossing and the effective potential switch. As anticipated in section \ref{sec:tamingthedemon}, this delay creates an effective hysteresis $h=|v| \tau$, where $v$ is the velocity of the oscillator when crossing the barrier. The absolute value in this expression stems from the fact that only the velocity sign that matches the barrier crossing is considered (for example positive velocity for upward crossing). The mean value $\langle h \rangle$ can be computed knowing the speed PDF: 
\begin{align}
\langle h \rangle & =\langle |v|\rangle\,\tau = \int_0^{\infty} |v| \frac{e^{-\frac{v^2}{2 \sigma_v^2}}}{\sigma_v \sqrt{2\pi}}dv\,\tau \\
&=\sqrt{\frac{2}{\pi}}\sigma_v \tau =\sqrt{\frac{2}{\pi}\frac{k_B T}{m}} \tau 
\end{align}
The switching rate $\Gamma(\beta \mathcal{B})$ between the 2 wells is given by Eq.~\ref{EqExtKramers}, and each crossing implies an energy extraction $\Delta U = 2 k x_1 h = 2 \sqrt{2k\mathcal{B}} h$. The mean power extracted from the oscillator is therefore
\begin{equation}
 \langle \dot U \rangle = \Gamma \langle \Delta U \rangle= \frac{4}{\sqrt{\pi}} k_B T \sqrt{\frac{\mathcal{B}}{k_B T}}\Gamma\left(\frac{\mathcal{B}}{k_B T}\right) \omega_0 \tau.
\end{equation}
In the steady state, this power is equilibrated by the heat flux coming from the bath, which is expressed within the framework of stochastic thermodynamics as~\cite{Dago-2022-JStat,Dago-2022-PRL}
\begin{equation}
\langle \dot{\mathcal{Q}} \rangle =\frac{\omega_0}{Q}k_B (T_0 -T) \label{eqbalancesimp} 
\end{equation}
Writing the out-of-equilibrium steady state balance $\langle \dot U \rangle = \langle \dot{\mathcal{Q}} \rangle$, we derive the following equation:
\begin{align}
\gamma\left(\frac{\mathcal{B}}{k_B T_0}\frac{T_0}{T}\right) Q \omega_0 \tau =\frac{T_0}{T} - 1 \label{eqTdelay}
\end{align}
where
\begin{align} 
\gamma(b)=\frac{4}{\sqrt{\pi}} \frac{\sqrt{b} e^{-b}}{2\pi - \pi e^{-b} + 2 \int_1^\infty d\epsilon b e^{-\epsilon b} \sin^{-1}({\epsilon^{-\frac{1}{2}}})}.
\end{align}
The solution to this equation are a function of $Q\omega_0\tau$, and can be numerically computed, and are plotted in Fig.~\ref{theta_map} versus the barrier height $\mathcal{B}$. As expected, the oscillator is warmed up by the demon when $\tau<0$, and cooled down when $\tau>0$. The amplitude of the temperature change depends on $\mathcal{B}$. For high $\mathcal{B}$, $\Delta U$ is large but barrier crossing are scarce, so the temperature is unchanged. On the contrary, for low barrier heights, crossings are frequent but imply a low energy exchange, leaving the system at the thermostat temperature. For intermediate values $B\sim k_BT_0$, crossing are frequent and induce a sizable energy exchange, thus an effective temperature departing from the $T_0$. 

The function $\gamma(b)$ presents a global maximum $\gamma^*=0.19$ in $b^*=0.35$, allowing to compute the extremum temperature and corresponding energy barrier
\begin{eqnarray}
T^* &=& \frac{T_0}{1+\gamma^*Q\omega_0\tau}\label{eq:thetamin}\\
\mathcal{B}^* &=& b^* k_B T^* \label{EqT*}
\end{eqnarray}
The extremum temperature is thus a function of $Q\omega_0\tau$. In particular, no kinetic temperature change is expected if $\gamma^* Q\omega_0\tau\ll 1$. Fig.~\ref{theta_map}(b) demonstrate how this description is pertinent and matches the experimental observations.

Taming the demon thus requires to be able tune the feedback delay to be much smaller than $1/(\gamma^* Q\omega_0)$, a challenge increasing with the quality factor of the oscillator. In our setup ($f_0=\SI{1270}{Hz}$), a quality factor $Q=100$ requires $\tau\ll\SI{6.6}{\mu s}$. With an deterministic delay of $\tau_d=\sim\SI{1}{\mu s}$ on top of the phase delay introduced by the filters, the feedback loop delay necessitate an anticipation of the signal to be transparent in such stringent conditions. To have an effective criterion on the phase of the signal conditioning process, let us fix a maximum heating/cooling to $5\%$ of $T_0$, that is $\gamma^* Q\omega_0|\tau|<0.05$. A time delay $\tau$ translates into a phase $\phi=\omega\tau$, hence the criterion is simply:
\begin{equation} \label{EqPhaseCriterium}
\phi<\frac{0.26}{Q}\frac{\omega}{\omega_0}.
\end{equation}
This criterion is reported in Fig.~\ref{FigFilters}. We see that we manage to fulfill the criterion in a reasonable frequency range around the resonance, where most of energy of the motion is concentrated. All in all, a well tuned feedback allows the kinetic temperature to match the thermostat one within 3\%. Moreover, the demon can also be unleashed on purpose by tuning the anticipation delay $\tau_a$ so as to create a non-equilibrium exchange of energy with the thermostat.

\section{Conclusions}

In this chapter, we describe how a virtual potential can be realized using a feedback loop to modify the dynamics of an underdamped system, in our case is a micro-mechanical oscillator. This technique produces a tunable non linear virtual potential which is indistinguishable from a physical one on the example demonstrated here: a bi-quadratic double well. To get such a result we take into account and overcome the drawbacks of the method, which are induced for example by the feedback delay and by noise in the commutation threshold between the two wells. We discuss in details how to fine tune the feedback parameters in order to get a ``perfect'' virtual-potential and we show how the delay (anticipation) of the feedback introduces a cooling (heating) of the device. As an example of application, we present a detailed experimental and theoretical analysis of the jumping rate between two potential wells showing the correction to Kramer's rate for small energy barriers.

The use of non linear virtual-potential in underdamped oscillators open the way to many new and interesting applications which present different features from those performed in overdamped systems. For example, our device has been recently used to measure Landauer's bound in underdamped systems\cite{Dago-2022-JStat,Dago-2021,Dago-2022-PRL,Dago-2023-PNAS,Dago-2023-APR}, underlying their advantages with respect to overdamped ones. Among other applications is the analysis of the energetics of the feedback: the device can be used as an information engine in which the information acquired by the feedback (the demon) can be used to produce work. We overlooked this aspect in this chapter, only showing that demons could be used to create non-equilibrium steady states with heat fluxes between the oscillator and the thermostat. The analysis of the demon from the information prism is still under study and will be the subject of future works.

\begin{acknowledgement}
This work has been financially supported by the Agence Nationale de la Recherche through grant ANR-18-CE30-0013 and by the FQXi Foundation, Grant No. FQXi-IAF19-05, ``Information as a fuel in colloids and super-conducting quantum circuits''.
\end{acknowledgement}

\ethics{Competing Interests}{The authors have no conflicts of interest to declare that are relevant to the content of this chapter.}

\nocite{apsrev41Control}

\bibliographystyle{apsrev4-2}
\bibliography{FeedbackPotential.bib}

\end{document}